\documentclass[final,5p,times,twocolumn]{elsarticle}

\usepackage[T1]{fontenc} 
\usepackage{lmodern}  
\usepackage[utf8]{inputenc} 
\usepackage{hyperref} 
\usepackage{graphicx} 
\usepackage[numbers]{natbib} 
\usepackage{amsmath}
\DeclareMathOperator{\MIP}{MIP}

\DeclareMathOperator{\erf}{erf}

\title{ILD Silicon Tungsten Electromagnetic Calorimeter First Full Scale Electronic Prototype}
\author{Frédéric Magniette, Jérôme Nanni, Rémi Guillaumat, Marc Louzir, Marc Anduze, Evelyne Edy, Oleksandr Korostyshevskyi, Vladislav Balagura, Vincent Boudry, Jean-Claude Brient \\on behalf of the ILD concept group\\
\footnotesize Laboratoire Leprince-Ringuet (LLR), \'Ecole polytechnique, CNRS/IN2P3, F-91128 Palaiseau, France}
\begin{document}
\maketitle

\begin{abstract}
The "long slab" is a new prototype for the SiW-Ecal, a silicon tungsten electromagnetic calorimeter for the International Large Detector (ILD) at the future International Linear Collider (ILC). The new prototype has been evaluated with cosmics, radioactive sources and with 3 GeV electrons in beam tests at the DESY facility, Hamburg. A channel-wise calibration has been achieved, at different angles of incidence of the beam on the sensors. Using data collected at non-normal incidence, the signal of particles traversing adjacent pixels were used to estimate the absolute value of the trigger threshold in units of mips. This new prototype provides us many hints on how to improve the design of the front-end electronics. It is also a convenient tool to estimate the critical characteristics of ILD SiW-Ecal (like power consumption, cooling, readout time, etc.) and to optimise the future design of the detector.
\end{abstract}

\smallskip
\paragraph*{Introduction}

\begin{figure}[h]
  \centering
  \includegraphics[width=0.4\textwidth]{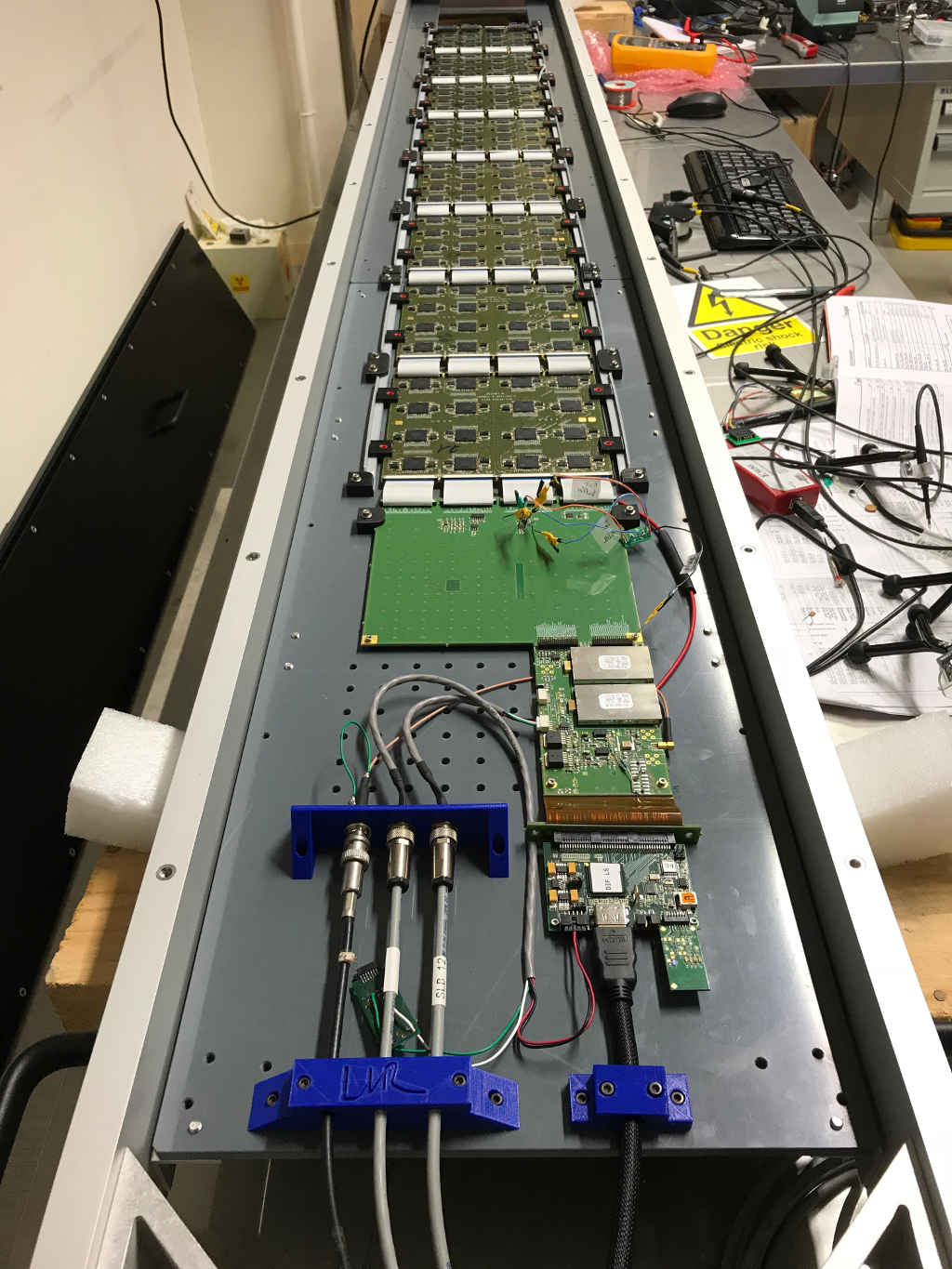}
  \caption{Assembled long slab with 8 ASUs.}
  \label{fig:sl_elec}
\end{figure}

The SiW-Ecal \cite{siwecal} is a sampling electromagnetic calorimeter developed for the ILD detector \cite{ild} of the future International Linear Collider (ILC). Its detector elements are called slabs and are based on high-resistivity silicon diodes of several hundred microns thickness, divided in pixels of $5 \times 5 \, mm^2$, alternated with tungsten absorbers. The previous prototypes tested a tower of ``short slabs'' featuring an active region of $18 \times 18 \, cm^2$ for each single ASU, and fully described in \cite{perf_short}. The long slab is a new prototype that has been designed to demonstrate that it is electronically viable to operate a long electronic board chain. It consists of 8 electronic front-end boards (named ASU for Active Sensor Unit), assembled together to create a $144 \times 18 \, cm^2$ detector which is the typical size of the slabs in the ILD barrel (Figure \ref{fig:sl_elec}). 

For this prototype, we have relaxed the geometrical constraints imposed by the ILD to focus on the electronics and the performance of the detector along the length. Each board has been equipped with a $320 \, \mu m$ thick small sensor ($2 \times 2 \, cm^2$ for $4 \times 4$ pixels) produced by the Hamamatsu Photonics company. 

\smallskip
\paragraph*{Mechanics}

\begin{figure}[h]
  \centering
  \includegraphics[width=0.4\textwidth]{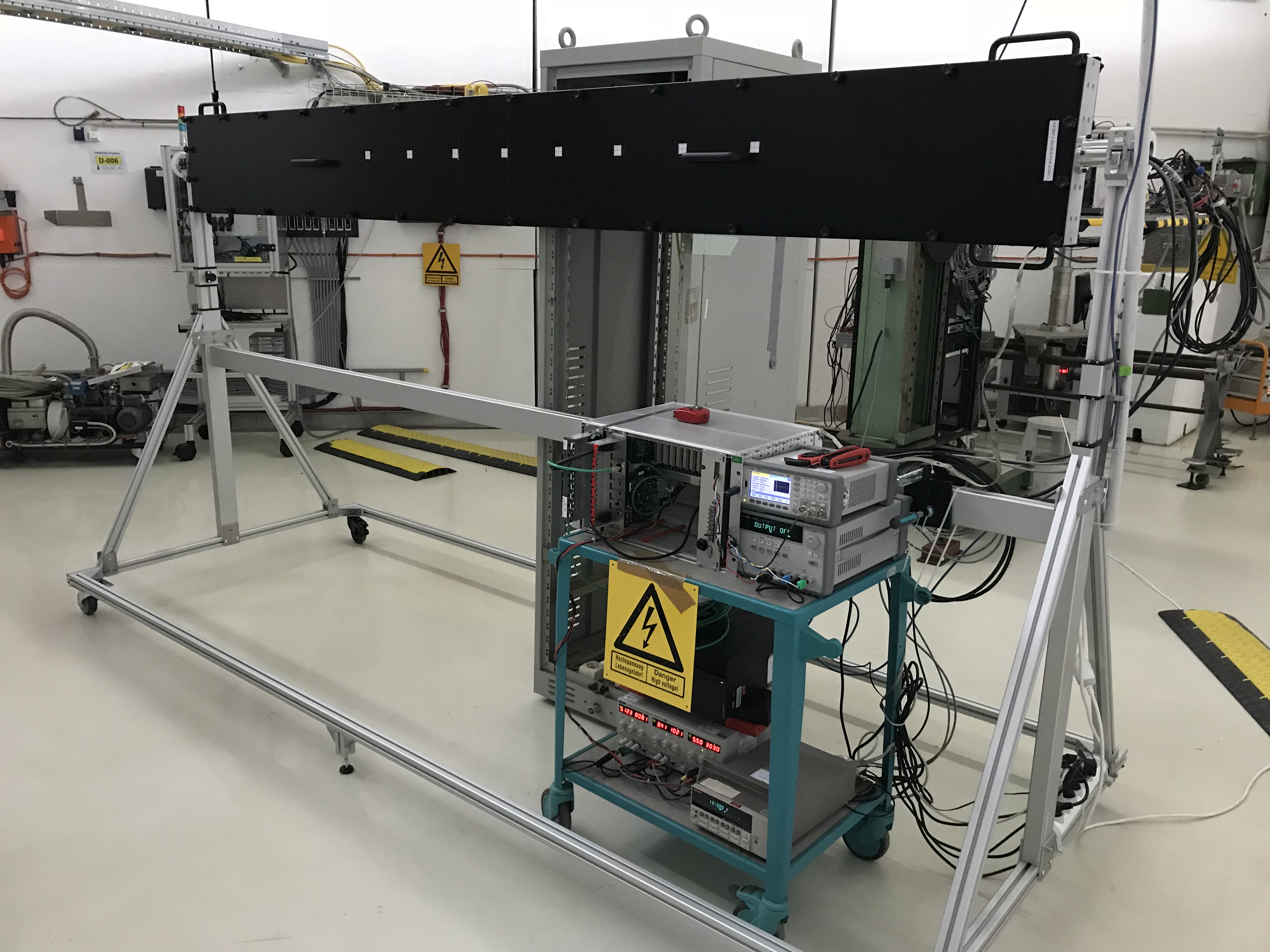}
  \caption{Long slab setup at DESY}
  \label{fig:sl_meca}
\end{figure}

For use in beam tests, a mechanical structure of 3 meters long, shown in Figure \ref{fig:sl_meca}, has been built, allowing to support the slab and to incline it in the beam with a precision of $1^\circ$. It ensures the mechanical rigidity (max 1 mm of bend over a 3 m length) to ensure reliable operation in all conditions. Shielding is added to avoid electro-magnetic induced noise. The directional blockable wheels allow placing the detector in the beam easily.

\smallskip
\paragraph*{Electronics}

The building block of the detector is the ASU. It is a multi-layered PCB with two very different external sides. The lower side is covered with pads with the size of the silicon pixels. The silicon sensors are glued on this side with a conductive glue. On the upper side sits the reading ASICs, named Skiroc 2 and designed by OMEGA group \cite{skiroc2}. The ASUs are chainable to obtain long slabs with an almost continuous active surface. At the end of the chain, a front-end board reads the ASICs and transmits data to the Data Acquisition System (DAQ) described in \cite{TIPP14}. The long slab prototype is the first attempt to chain the ASUs to test their communication and performance. For testing purpose, flat cables have been chosen for data and clocks to ease dismounting without degrading transmission performance. The impedances and adaptive systems (called drivers) have been tuned for every differential line by termination resistors. For bias voltage, shielded cables has been used with lemo connectors, allowing to reduce the electromagnetic noise and ease the connection. 

The length of the prototype induces difficulties for clock and signal propagation and data integrity. In particular, reflections appear on the clock lines. Indeed, the clock line has a comb shape (in order to be evenly distributed), and the signal bounces in the stubs, creating reflection. As a consequence, the clock signal is not square but include glitches in the transition region. This creates extra clock beats and does not allow to transfer the configuration bitstream correctly to the ASICs. This problem was solved by adjusting the level of the glitches by adding an RC filter in front. The filter parameters were optimized by simulating an isolated clock line using the Cadence Sigrity tool \cite{sigrity}, integrating the PCB electrical specification and the driver characteristics. Even if the RC filter permits the use of the actual prototype, a new design of clock lines has to be developed in future versions of ASU to avoid the comb shape, as it has been done for data lines.

Another difficulty induced by the length of the prototype is the increased level of noise on the sensor's bias voltage. The SiW-Ecal readout is very sensitive to this noise which must be reduced to the minimum. To achieve this goal, we insert RC filters on the bias line between the ASUs. We use third order filters, adapted to reduce high frequency noises above 100 kHz. In the final version, for compacity, it is planned to use a Kapton sheet instead of cables; a change of the design will be required to include these RC filters.

Finally, it has been observed that the bandwidth of the DAQ is not entirely sufficient when the long slab has all its memories full. For example, in case of a pedestal calibration, it is necessary to read all channels of all chips at the same time. In the future, we need to use a parallel reading of partitions of ASICs inside the front-end board. 

\smallskip
\paragraph*{Testing}

The long slab performance has been tested with cosmics, radioactive sources and with 3 GeV electrons in the beam tests at DESY, Hamburg. In particular, radioactive sources have been used at every step of mounting a new ASU to test the behaviour of the new board and also to check the noise induced by the card on the others. This methodology allowed us to detect the problems before they mingle with another one. 

During the test beam at DESY, we have accumulated data for both normal and inclined incidence of the beam to the silicon sensor surface. The positioning of the mechanical structure was performed with a laser to ensure full exposure of every pixel. We took data in all the pixels at three different angles of incidence to the silicon sensor surface (0, 45 and 60 degrees).

\smallskip
\paragraph*{MIP fit}

The Minimum Ionizing Particle (MIP) energy deposition is determined by fitting the pedestal subtracted  analog-to-digital converter (ADC) spectra. This pedestal varies among the ASICs memories and channels and thus has to be evaluated for every memory associated with every channel. Once the pedestal subtracted, we fit the histogram of the energy deposition in every pixel. This is fitted by a sum of two Landau distributions convolved with a Gaussian. The first Landau distribution corresponds to a single MIP hitting the pixel during the integration time. The second Landau distribution models 2 MIPs (pile-up). The contribution of a third Landau distribution can be neglected.

\begin{figure}[h]
  \centering
  \includegraphics[width=0.4\textwidth]{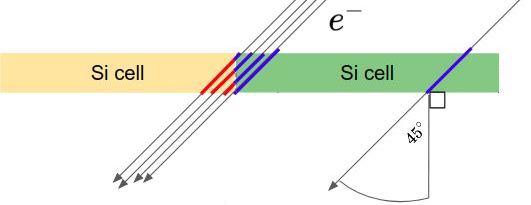}
  \caption{When the prototype is inclined, the incident particle can cross two pixels, modifying the energy deposition spectrum.}
  \label{fig:ang}
\end{figure}

\begin{figure}[h]
  \centering
  \includegraphics[width=0.49\textwidth]{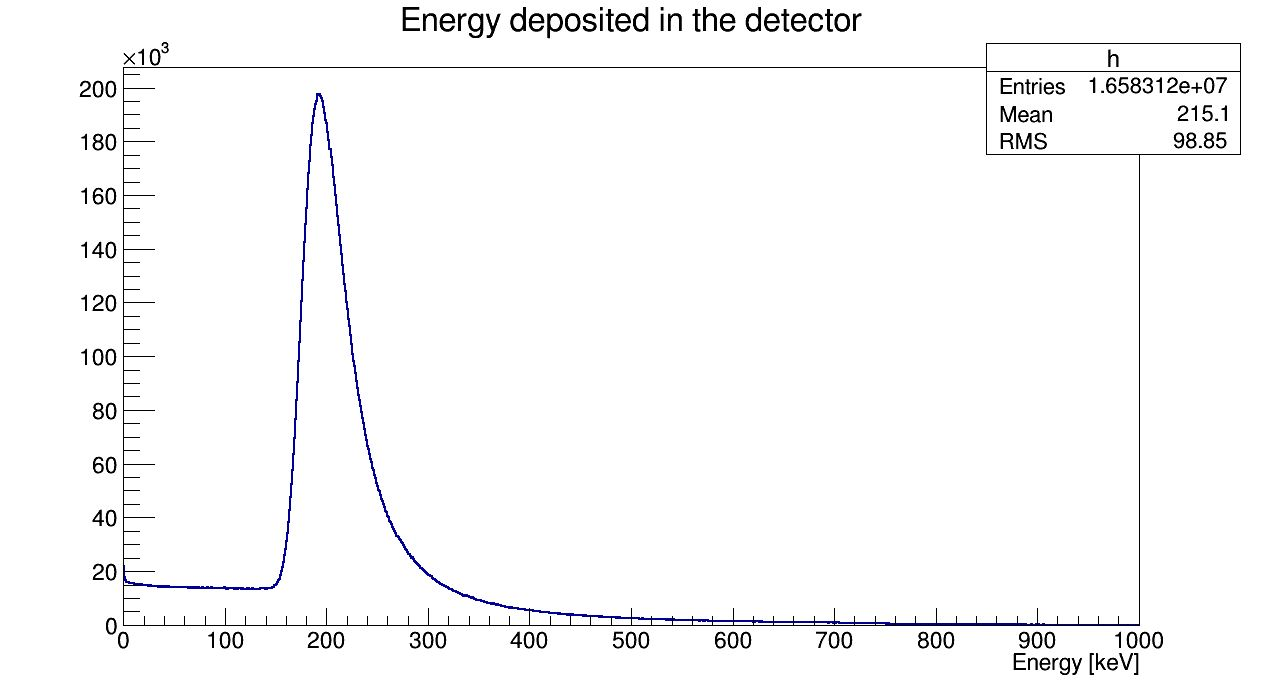}
  \hfill
  \caption{On the GEANT4 simulation of 2 regions, the two-pixels energy sharing adds a low energy plateau to the single-pixel energy deposition histogram.}
  \label{fig:simul}
\end{figure}

As shown in Figure \ref{fig:ang}, with the inclined beam, the particle can sometimes go across two pixels. For a sufficiently uniform beam and in the simplest model when electron-hole pairs in the  silicon always drift perpendicularly to its surface, the one-pixel signal is reduced in this case such that it has equal probabilities of being anywhere between zero and the total energy deposition in the silicon shared by two pixels. The energy deposited distribution is then modified. In particular, it starts not from zero on the left but from a plateau which remains flat until the point where the original Landau distribution starts growing significantly. A simulation of this situation has been performed with Geant4, presented in Figure \ref{fig:simul}, showing clearly the plateau on the left of the energy deposition histogram.

This plateau and the left part of the Landau distribution can be modelled by assuming a uniform energy sharing of the deposited energy between two pixels. Let's denote dE/dx Landau distribution by $L(x)$ and the probability that the particle passes though two pixels by $c$. Then, the normalized distribution of energies  deposited in one pixel is:

\begin{equation}
f(x)=(1-c)\frac{L(x)}{\int_{0}^{+\infty}L(t) \, dt}+c\frac{\int_{x}^{+\infty}\frac{L(t)}{t} \, dt}{\int_{0}^{+\infty}\int_{x}^{+\infty} \frac{L(t)}{t} \, dt \, dx}
\label{equ:modang}
\end{equation}

where the second term describes the energy sharing. Namely, the total energy $L(t)$ in the enumerator $\int_x^{+\infty} L(t)/t\; dt$ is uniformly distributed in the range $[0 ... L(t)]$ and creates the density $L(t)/t$ which is then integrated over $L(t)$ for all $t \geq x$. This plateau is ideally suited for measuring the position and the shape of the trigger threshold, modelled by an error function denoted in the following by erf. The full fit can be performed with the following function:

\begin{equation}
\frac{\erf(x)+1}{2} \int_{-\infty }^{+\infty}f(t) \cdot \mathrm{Gauss}(x-t,\sigma)dt
\label{equ:fullfit}
\end{equation}

where $\mathrm{Gauss}(x, \sigma)$ denotes the Gaussian with the center $x$ and the sigma $\sigma$.

\begin{figure}[h]
  \centering
  \includegraphics[width=0.45\textwidth]{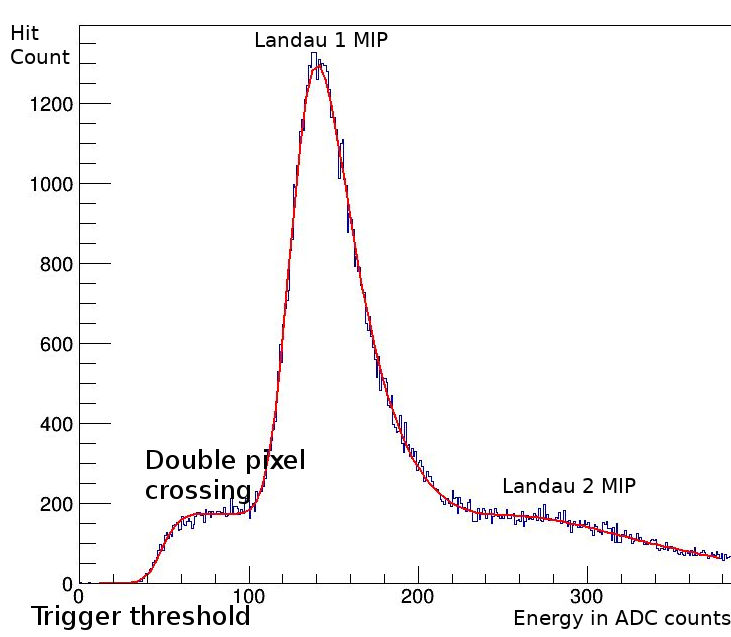}
  \caption{Full fit of the pedestal-subtracted signal histogram : two Landau distributions representing 1 and 2 MIPs respectively are convolved with a Gaussian. On the left, the pixel-sharing plateau is truncated by an erf function.}
  \label{fig:mipfit}
\end{figure}

The full fit is shown in Figure \ref{fig:mipfit}. The Most Probable Value (MPV) of the first Landau distribution gives the value of the MIP in ADC counts. The inflexion point of the erf function provides the value of the trigger threshold. Figure \ref{fig:mipfit}, therefore clearly demonstrates that the threshold was 
set to sufficiently low value far below the  MIP signal. With these two pieces of information, we can infer the value of this threshold in term of a MIP percentage which is very valuable to perform accurate calibration of the system.

\smallskip
\paragraph*{MIP uniformity along the slab}

\begin{figure}[h]
  \centering
  \includegraphics[width=0.5\textwidth]{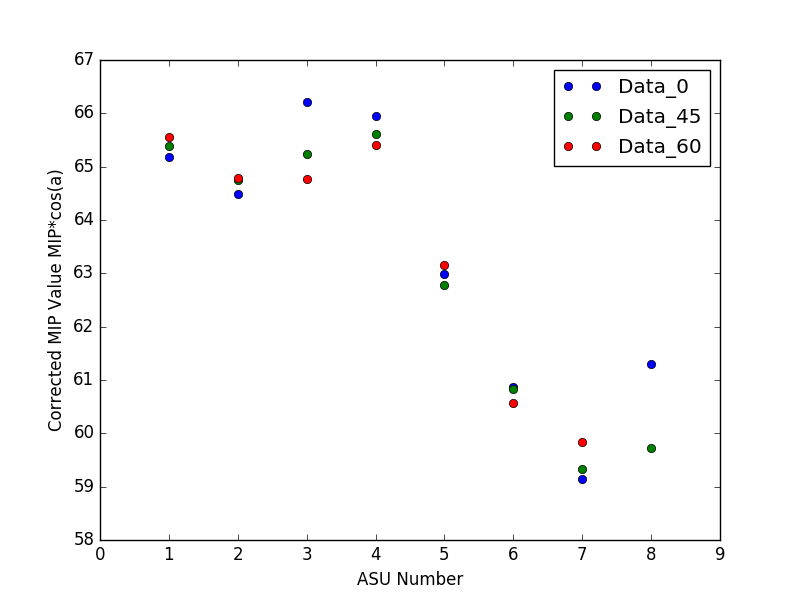}
  \caption{MIP values corrected for incidence angle for all the ASUs of the prototype, at the incidence angles (0, 45 and 60 degrees)}
  \label{fig:allmips}
\end{figure}

With this MIP fit function, we can get accurate values of the MIP on all the sensors, depending on the angle. Figure \ref{fig:allmips} summarises the extracted MIP values for the 8 ASUs for the three considered angles. A $1/\cos(\alpha)$ relates the three series as expected from the proportionality of the MIP value and the thickness of silicon crossed by the particle. On Figure \ref{fig:allmips}, the MIP values are corrected. Points corresponding to the last ASU under 45 and 60-degree angle are missing because the corresponding sensor was inaccessible due to the size of the beam room.

\begin{figure}[h]
  \centering
  \includegraphics[width=0.49\textwidth]{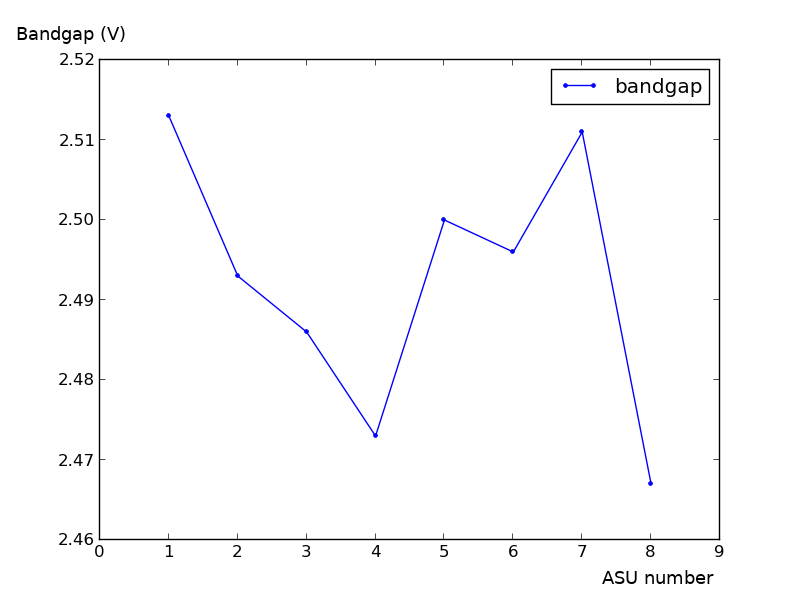}
  \caption{The measured bandgap of the 8 ASIC of the long slab.}
  \label{fig:bandgap}
\end{figure}

In an ideal prototype, the three series should be completely flat but they are not. They show a strange shape including a slope of about 10 \% in the value of the MIP. This kind of variations may come from different sources. We have investigated the ASIC's power supply voltage along the length of the prototype and measured a 150 mV linear decrease over the total length. This might explain a part of the variation. We have also investigated a reference voltage, so-called bandgap, an internal reference inside the readout ASIC which can change the amplification and thus the MIP value. Figure \ref{fig:bandgap} shows the variation of this reference on the different ASU.

A more detailed measurement on all 128 ASICs has shown that this bandgap varies from ASIC to ASIC. There is no systematic effect and the distribution of the bandgap over the ASICs seems to follow a normal distribution with a standard deviation of 19.2 mV and a maximum peak-to-peak difference of 200 mV.

\begin{figure}[h]
  \centering
  \includegraphics[width=0.49\textwidth]{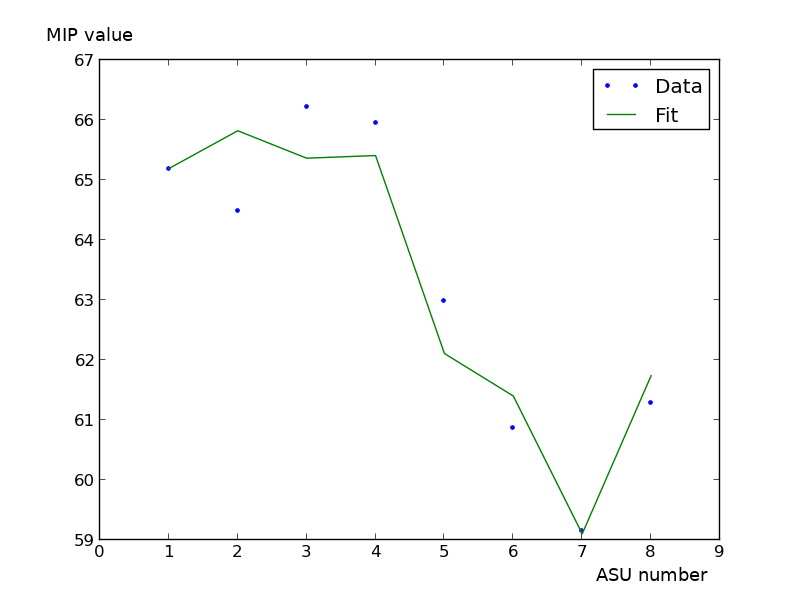}
  \caption{Fit of the MIP values along the length of the prototype (in ASU)}
  \label{fig:fitgap}
\end{figure}

Based on the measured ASIC power and bandgap voltages, an attempt has been made to reproduce the MIP variation across ASUs using a simple model of the form:
\begin{equation}
\MIP(\mathrm{ASU})=a*\mathrm{ASU}+b-c*\mathrm{bandgap(ASU)}.
\end{equation}
The best fit on 0 degree incidence MIP values is shown in Figure \ref{fig:fitgap}. Further studies are needed to understand fully the observed variation of the MIP signal across ASUs.

\smallskip
\paragraph*{Conclusion and Perspectives}

The electronic long slab prototype has proved that the ASUs are almost suitable for chained operation. After solving the different electronics problems, we found out that the long slab has almost the same performance at low energy (MIP) as the short slabs tested previously \cite{perf_short}.

Nevertheless, many improvements could be envisaged for the next version of the ASU:
\begin{itemize}
\item Clock line adaptation: the shape of this line has to be optimised to avoid reflections and limit the possibility of bitstream corruption
\item RC filtering of the sensor's bias voltage: the current design uses a long Kapton covering all ASUs. It should be replaced by a single Kapton per ASU design to integrate a low-noise filter.
\item The power supply of the ASICs has to be stabilised along the length of the slab, for example by equalising the length of all the power lines.
\item The analysis has proved that the bandgap disparity between ASICs is not neglectable. Improvement requires the refactoring of the ASIC design (which has already been partly done with the new version Skiroc 2a). In case the bandgap spread between ASICs remains too large, they will have to be individually measured and compensated by software.
\item The sequential reading of the ASIC limits the bandwidth and can cause problems during pedestal or charge injection calibrations. A solution with readout partitions has to be considered.
\end{itemize}

The successful operation of a long slab is a milestone toward a proof of the feasibility of the ILD SiW-Ecal. A prototype including all the previously mentioned improvements, the use of broader and thicker sensors and including all the mechanical constraints of the ILD experiment still needs to be built and tested with high energy electrons to fully assess the design.

\smallskip
\paragraph*{References}

\bibliographystyle{unsrt} 
\bibliography{vci19}

This project has received funding from the European Union’s Horizon 2020 Research and Innovation programme under Grant Agreement no. 654168. The P2IO labex supported it under the HiGHTEC project.

\end{document}